\documentclass{llncs}
\usepackage{hyphenat}
\usepackage{makeidx}
\usepackage{todonotes}
\usepackage[utf8]{inputenc}
\usepackage{listings}
\usepackage[export]{adjustbox}
\usepackage{float}
\usepackage{url}
\usepackage{hyperref}
\usepackage{pgf-pie}

\usepackage{soul}
\usepackage{tikz}
\usetikzlibrary{calc}

\makeatletter

\newif\if@anonymize

\@anonymizefalse  

\if@anonymize
  \newcommand{\highlight@DoHighlight}{
    \fill [outer sep = -15pt, inner sep = 0pt, color=black]
          ($(begin highlight)+(0,8pt)$) rectangle ($(end highlight)+(0,-3pt)$) ;
  }

  \newcommand{\highlight@BeginHighlight}{
    \coordinate (begin highlight) at (0,0) ;
  }

  \newcommand{\highlight@EndHighlight}{
    \coordinate (end highlight) at (0,0) ;
  }

  \newdimen\highlight@previous
  \newdimen\highlight@current
  \newlength{\item@width}

  \DeclareRobustCommand*\anonymize{%
    \SOUL@setup
    \def\SOUL@preamble{%
      \begin{tikzpicture}[overlay, remember picture]
        \highlight@BeginHighlight
        \highlight@EndHighlight
      \end{tikzpicture}%
    }%
    \def\SOUL@postamble{%
      \begin{tikzpicture}[overlay, remember picture]
        \highlight@EndHighlight
        \highlight@DoHighlight
      \end{tikzpicture}%
    }%
    \def\SOUL@everyhyphen{%
      \discretionary{%
        \SOUL@setkern\SOUL@hyphkern
        \SOUL@sethyphenchar
        \tikz[overlay, remember picture] \highlight@EndHighlight ;%
      }{%
      }{%
        \SOUL@setkern\SOUL@charkern
      }%
    }%
    \def\SOUL@everyexhyphen##1{%
      \SOUL@setkern\SOUL@hyphkern
      \settowidth{\item@width}{##1}%
      \makebox[\item@width]{}%
      \discretionary{%
        \tikz[overlay, remember picture] \highlight@EndHighlight ;%
      }{%
      }{%
        \SOUL@setkern\SOUL@charkern
      }%
    }%
    \def\SOUL@everysyllable{%
      \begin{tikzpicture}[overlay, remember picture]
        \path let \p0 = (begin highlight), \p1 = (0,0) in \pgfextra
          \global\highlight@previous=\y0
          \global\highlight@current =\y1
        \endpgfextra (0,0) ;
        \ifdim\highlight@current < \highlight@previous
          \highlight@DoHighlight
          \highlight@BeginHighlight
        \fi
      \end{tikzpicture}%
      \settowidth{\item@width}{\the\SOUL@syllable}%
      \makebox[\item@width]{}%
      \tikz[overlay, remember picture] \highlight@EndHighlight ;%
    }%
    \SOUL@
  }
\else
  \newcommand{\anonymize}[1]{#1}
\fi
\makeatother

\begin{document}

\mainmatter
\title{The Open Service Compendium}
\subtitle{Business-pertinent Cloud Service\\Discovery, Assessment, and Selection.}

\titlerunning{Running Title}

\author{Mathias Slawik \and Begüm İlke Zilci \and Fabian Knaack \and Axel Küpper}

\authorrunning{Slawik, et al.}

\institute{Telekom Innovation Laboratories, Technische Universität Berlin\\Service-centric Networking\\\email{mathias.slawik|ilke.zilci|fabian.knaack|axel.kuepper@tu-berlin.de}}

\maketitle

\begin{abstract}
When trying to discover, assess, and select cloud services, companies face many challenges, such as fast-moving markets, vast numbers of offerings, and highly ambiguous selection criteria. This publication presents the Open Service Compendium (OSC), an information system which supports businesses in their discovery, assessment and cloud service selection by offering a simple dynamic service description language, business-pertinent vocabularies, as well as matchmaking functionality. It contributes to the state of the art by offering a more practical, mature, simple, and usable approach than related works.

\end{abstract}

\section{Introduction}

There is a major trend within enterprise IT to fundamentally embrace cloud computing. The most recent 2015 "State of the Cloud Survey" reveals that 93 percent of large enterprises (i.e. 1000+ employees) are already using cloud computing solutions, 82 percent follow a multi-cloud strategy, while only 3 percent do not have plans for adopting cloud computing\footnote{\url{http://goo.gl/eloh66}}.

Before companies contract and consume cloud services, they have to carry out \textit{discovery}, i.e., finding cloud services in the vast Internet, \textit{assessment}, i.e., matching services to requirements, and \textit{selection}, i.e, choosing the best service for subsequent booking and consumption, e.g., by making a shortlist and ranking services. These tasks are challenging: cloud markets are fast-moving, have a vast numbers of offerings, selection criteria are highly ambiguous, marketplaces sometimes unorganized, and price structures and feature combinations complex and opaque. These challenges impede optimal service selection and sometimes hinders cloud adoption generally.

Our contribution was conceived within two research projects targeting specific domains: TRESOR\footnote{\url{http://www.cloud-tresor.com}} targeting the German Health sector and CYCLONE\footnote{\url{http://www.cyclone-project.eu}} targeting users of federated, multi-cloud applications. TRESOR developed a cloud ecosystem featuring a cloud broker \& marketplace and was thoroughly presented in our previous works \cite{SZT,TSZ,T12,ZST14}. CYCLONE is a Horizon 2020 innovation action which aims at integrating existing cloud management software to allow unified management of federated clouds. Both projects also address discovery, assessment, and selection challenges due to the lack of a suitable information systems: The TRESOR health centers cannot assess legal cloud consumption prerequisites, e.g., how and where medical data is processed. This leads to higher costs for local IT infrastructure and less functionality available to personnel. Many of the CYCLONE multi-cloud application developers face challenges in selecting optimal offerings for use in their applications, e.g., IaaS VMs and storage services. Suboptimal offerings can cause higher costs as well as lower Quality of Experience by the end-users of such applications.

Our previous work \cite{SK14} establishes basic technologies for addressing these issues: a textual domain specific language for describing services, a pertinent business vocabulary of selection criteria, and a brokering component. The analysis of related work showed a particular lack of pertinent service selection criteria in description languages as well as contemporary and future marketplaces, although there has been extensive empiric research in this area. Also, the benefits of using textual domain-specific languages \cite{F11} is not utilized in any of the examined approaches, which predominantly use semantic technologies for capturing service information. The main contribution of this publication is the design, implementation, and evaluation of a first iteration of the OSC, which is an information system supporting business users in their cloud service discovery, assessment, and selection activities. For this, we employ and extend former contributions and address the following research questions in this publication:

\begin{itemize}
\item[Q1] What are the main business challenges the OSC has to address and where should it differentiate itself from the state-of-the-art and related works?
\item[Q2] Which use cases should the OSC architecture implement and how should it be designed to be suitable, scalable, and state-of-the-art?
\item[Q3] How does the current OSC implementation meet its requirements, as well as the needs of its first users?
\end{itemize}

By addressing these research questions we extend existing research on description languages, matchmakers, and marketplaces, by including real-world requirements to further maturate this area of research. By designing the system to be "wiki-like" and having it used by regular Internet users we hope to increase the volume of empiric knowledge.

We apply the Information Systems Research Framework by Hevner et al. (\cite{HMPR04}) which also structures this publication: Chapter \ref{sec:business-challenges} iterates prevalent business challenges and derives eight main OSC requirements. These requirements are contrasted with the related work in Chapter \ref{sec:related-work} to guide the OSC use case definition in Chapter \ref{sec:osc-usecases}. Based on these use cases we design the OSC architecture in Chapter \ref{sec:architecture} and present its current implementation status. After showing first evaluation results in Chapter \ref{sec:evaluation}, we conclude this publication in Chapter \ref{sec:conclusion}.

\section{Cloud Service business challenges}
\label{sec:business-challenges}

This chapter structures the discovery, assessment, and selection challenges into three problem areas, describes them, and specifies requirements for the OSC, numbered \textbf{R1} to \textbf{R8}.

\textbf{Cloud Market Characteristics: Fast-moving Vastness.} The cloud market is \textit{vast} and \textit{fast-moving}: Current forecasts demonstrate its increasing \textit{vastness}: the total end-user spending on public cloud services is expected to grow by almost 60\% between 2015 and 2018 to a staggering \$290bn\footnote{\url{http://www.ft.com/cms/s/2/b3d40e7a-ceea-11e3-ac8d-00144feabdc0.html}}. Some cloud vendors are also astonishingly large: Amazon Web Services, for example, has more than 1 million customers, achieved more than 40 percent year-over-year revenue growth, and generates an estimated yearly revenue of \$4 billion \footnote{\url{http://goo.gl/5vHSom}}. The "Google Memorial"\footnote{\url{http://goo.gl/YdN2np}} highlights the velocity of a \textit{fast-moving} cloud market participant: it lists 66 discontinued services which were sometimes highly popular, for example, the Google Reader service had more than 24 million users\footnote{\url{http://googlesystem.blogspot.de/2013/03/google-reader-data-points.html}} before it was suddenly discontinued in 2013. These examples highlight that the cloud market is too vast for companies to obtain an optimal overview and it is too fast-moving to keep up with ever-changing service offerings. These cloud market characteristics require a \textit{structured service repository} \textbf{(R1)}, which integrates \textit{dynamic information} \textbf{(R2)}, e.g., IaaS spot-market prices. For maximum impact, it should be \textit{"wiki-like"}, i.e., any Internet user should be able to create and edit service descriptions \textbf{(R3)}. A \textit{matchmaking} between requirements and contained knowledge should support service selection \textbf{(R4)}.

\textbf{Ambiguous criteria and scattered knowledge.} Assessing service offerings raises two questions: \textit{what criteria to use} and \textit{where to get the required information}. Deciding what criteria to use is hard: they are sometimes highly ambiguous (e.g., data privacy criteria as shown by Selzer \cite{S14}) and sometimes empirically identified, yet neither integrated into service description languages, nor existing marketplaces and repositories, as we'll show in the next chapter. Gathering information to apply these criteria is also a challenging task: First of all, companies conceal knowledge about unfavorable service aspects. For example, cloud backup providers label services "unlimited", while they have in fact bandwidth and storage limits\footnote{\url{http://goo.gl/nVqeA3}}. The "Fair Use" clause of Backblaze, which allows the provider to cancel the contract anytime \footnote{\url{https://www.backblaze.com/terms.html}}, and the CrashPlan "Unlimited" limits which are concealed in the EULA \footnote{\url{http://support.code42.com/CrashPlan/CrashPlan\_For\_Home\_EULA}} highlight this practice. Secondly, some companies provide insufficient information: for example, Microsoft states that OneDrive can only be used with the Windows 8.1 Explorer if users use their live.com accounts for logging on to Windows \footnote{\url{http://windows.microsoft.com/en-us/windows-8/onedrive-app-faq}}.
On the contrary, many companies are reluctant to allow their corporate users to use such accounts, thus hindering them to use OneDrive effectively. Only private blogs offer workarounds
which are not always discovered by companies wishing to assess OneDrive \footnote{\url{http://goo.gl/PZ7d6p}}. In summary, as provider information does not suffice and knowledge becomes more scattered, the efforts for assessing services rise constantly unless there is a vocabulary of \textit{selection criteria pertinent to businesses} \textbf{(R5)}, as well as means of \textit{integrating external information} \textbf{(R6)}. 

\textbf{Features and Prices: Complex and Incomparable} In his seminal 1956 paper, Smith outlined that product differentiation and market segmentation are viable marketing strategies \cite{S56}. This observation still holds true almost sixty years later: to compete with cloud market leaders, service providers differentiate products and segment their market. One example is the online storage market, which is segmented into related categories, such as "remote backup", "cloud storage", and "file sharing". Different needs of consumers are addressed by different features and pricing schemes. "Cloud storage" services, such as Google Drive, allow flexible sharing of data, but incur additional costs for extending the free quota. "Backup services", such as CrashPlan, allow "unlimited" data storage for a fixed price but have only limited sharing functionality, e.g., backup family plans, such as "CrashPlan for Home". Thus, comparing different services becomes challenging if cloud consumers need to both share and backup large volumes of data. The price structure and feature combinations can also become complex: for example, Amazon EC2 offers 32 VM types in 10 locations with 6 operating systems, resulting in 1920 configuration options to choose from; in addition to opting for either on-demand, reserved, or spot market instances. Thus, comparing different competitors to find an optimal product implies an enormous effort, unless there is a suitable \textit{price model} \textbf{(R7)} as well as a mechanism for describing different \textit{variants} of a service \textbf{(R8)}.

\section{Related Work}
\label{sec:related-work}

The preceding business challenges are addressed by a number of related works from academia and practitioners in the areas of \textit{service description languages}, \textit{repositories and marketplaces}, \textit{service matchmaking approaches}, as well as \textit{cloud-selector frameworks}.

\textbf{Service Description Languages.} Dervice description languages capture elevant aspects of services for a specific use case. For example, WSDL\footnote{\url{http://www.w3.org/TR/wsdl20/}} and the CORBA IDL describe the technical service interface for the main use case of automated code generation of service skeletons. There is a wealth of service description approaches in the field of semantic web services, e.g., OWL-S \cite{MPMB+05}, WSMO \cite{WSMO}, SAWSDL\footnote{\url{http://www.w3.org/TR/2007/REC-sawsdl-20070828/}}, WSDL-S \footnote{\url{http://www.w3.org/Submission/WSDL-S}}, SWSF\footnote{\url{http://www.w3.org/Submission/SWSF/}}, and others. Other languages focus on business-related information, such as WSMO4IoS \cite{WSMO4IOS} as well as the Linked-USDL \cite{PCL14} which is derived from the earlier USDL \cite{OBKH13}. Other researchers, e.g., Breskovic, et al., create standardized products for electronic markets \cite{BAB13}, based on description languages, such as the CRDL \cite{RA10}. At last, some authors focus on price and cost modeling of cloud services \textbf{(R7)}, for example Kashef and Altmann in \cite{KA12} and \cite{AK14}. While these works provide interesting application areas for the OSC, they are not based on an existing service description language.

Seminal works by Fensel, et al. \cite{FFST11} as well as Studer, et al. \cite{SGA07} present semantic web services in detail. Studer, et al. summarizes the focus areas of semantic web services: reasoning-based matching of service functionality, harmonizing data formats and protocols, and automated Web Service composition. Semantic functionality requires service knowledge expression, e.g., service inputs and outputs, preconditions, and service effects. Therefore related languages have a broad scope and aims: for example, "maximise to the extent possible the level of automation" \cite{PCL14} or "covering as many XaaS domains as possible" \cite{SS13}. In contrast, cloud service discovery, assessment and selection activities by SMEs require merely a small set of relevant information, which has to be pertinent to business users. Yet, no approach meets the business challenges sufficiently: they do not handle \textit{dynamic information} \textbf{(R2)} well, cannot be used \textit{wiki-like} \textbf{(R3)}, nor can easily \textit{integrate external information} \textbf{(R6)}. While there are empiric studies on service selection criteria, e.g., \cite{RZWK12} and the CloudServiceCheck\footnote{\url{http://www.value4cloud.de/de/cloudservicecheck}}, the languages do not capture such service knowledge \textit{pertinent to businesses} \textbf{(R5)}. Another issue is the missing service variant management\footnote{The USDL "variant management" connotes variants \textit{of the language}.}. Without having a rich variant model, describing real-world cloud services becomes a major challenge, demonstrated by the example Amazon EC2 Linked-USDL description\footnote{\url{https://goo.gl/ZrMCWk}}. It only considers one type of instance and only one location, but consists of 1899 lines. We approximate a complete EC2 description to be 300.000 lines in length. This shows the prohibitive complexity of real-world USDL descriptions leading to inefficient processing and low human comprehensibility.

We argue that the failure to address existing SME challenges is the reason for its missing industry adoption outside of their funding scope. Many languages are abandoned, e.g., OWL-S (2006), SAWSDL (2007), WSMO (2008), and USDL (2011), and the associated tools are not updated anymore, e.g., the WSMO Studio\footnote{\url{http://sourceforge.net/projects/wsmostudio}}. Therefore we do not consider Semantic Web Service approaches suitable as the basis of the Open Service Compendium.

\textbf{Repositories and Marketplaces} address the vastness of the Cloud market by managing a large number of service descriptions. They can be divided into academic marketplace research platforms and high-volume SaaS marketplaces. There are academic marketplace research platforms which are relevant to our contribution: The USDL marketplace\footnote{\url{http://sourceforge.net/projects/usdlmarketplace}}, a proof-of-concept marketplace prototype based on USDL. The FI-Ware Marketplace and Repository\footnote{\url{https://github.com/service-business-framework}} provide APIs to manage USDL service descriptions, as well as support discovering and matching application and service offerings. At last, Spillner and Schill offer an extensible XaaS Service Registry which is based on WSMO4IoS \cite{SS13}.

We observe that no approach overcomes the explicated limitations of its SDL. Furthermore, none is designed to be used by regular Internet users, thus hindering their broad adoption and lowering their pertinence to businesses. Prototypical high-volume commercial SaaS marketplaces are the Google Marketplace \footnote{\url{https://www.google.com/enterprise/marketplace/home/apps}} and Salesforce AppExchange \footnote{\url{https://appexchange.salesforce.com}}. Instead of an elaborate cloud service formalization, they utilize very simple data models consisting of free-text, images, provider info, and a categorization. As they lack a service formalization, they are only marginally able to support users in their service discovery, selection, and assessment.

\textbf{Service Matchmaking.} The related work on service matchmaking is highly divergent in its contexts (e.g., Cloud Services, SOA, the Semantic Web), as well as on its opinions about what constitutes a matchmaking problem (e.g., the types of variables and the desired matchmaker functionality). We examine, if the existing service matchmakers are \textit{business-pertinent} \textbf{(R5)}. Our previous survey \cite{ZSK15b} divides approaches into \textit{syntactic}, \textit{constraint based}, \textit{ontological} and \textit{Fuzzy Set Theory based}: \textit{Syntactic} approaches are limited to numeric QoS parameters \cite{eleyan2011service,kritikos2009mixed}. \textit{Constraint based} transforming the service request into a set of constraints and match it to a set of service descriptions \cite{kritikos2009mixed}. Afterwards, the "closest" service can be found using the Euclidean distance between the request and the description \cite{eleyan2011service}. \textit{Ontological} approaches utilize OWL-S and reasoners, for example, to calculate the semantic similarity of method signatures \cite{liu2009weighted}, and to define the constraints as SWRL rules \cite{jie2011dynamic}. \textit{Fuzzy Set Theory} based approaches aim to match numeric QoS parameters in a flexible manner, sometimes extending syntactic approaches, for example, by allowing "good", "medium", and "poor" value intervals \cite{mobedpour2013user}, or using trapezoidal fuzzy numbers \cite{bacciu2010adaptive}.

While being extensive, none of the related approaches addresses current business challenges: most of them only consider numeric QoS parameters, such as availability, response time, and throughput \footnote{\url{http://www.uoguelph.ca/~qmahmoud/qws/}}, which are not independently and objectively measurable. Furthermore, pertinent selection criteria which are not numeric, e.g., feature lists, are not considered, as was shown in our previous publication \cite{ZSK15}. The referenced empiric studies show that only a small subset of relevant selection criteria are numeric. At last, none of these approaches uses more sophisticated description formats than WSDL.

\textbf{Cloud-selector frameworks} help cloud users in assessing different aspects of cloud providers. One example is PlanForCloud\footnote{\url{https://planforcloud.rightscale.com}} which allows users to create deployment descriptions and specify their planned usage of cloud resources, e.g., servers, storage, and databases. CloudHarmony\footnote{\url{https://cloudharmony.com}} is a bundle of services by Gartner: a provider directory, a benchmark database for network performance, and a service status dashboard. CloudSpectator\footnote{\url{http://cloudspectator.com}} offers performance measurements for different IaaS providers. Many limitations persist: PlanForCloud contains only services supported by RightScale software. CloudHarmony is quite extensive, yet lacks information about pricing and other business-pertinent selection criteria. None of the platforms offers matchmaking functionality or crowdsourcing data. The criteria are also limited, e.g., PlanForCloud offers only "price", while CloudSpectator supports only a "Multi Core Score".

\section{OSC Use-cases}
\label{sec:osc-usecases}

\begin{figure}
\vspace*{-0.5cm}
\centering
\includegraphics[width=8cm, clip=true, trim=0.25cm 22.5cm 5.4cm 0.3cm]{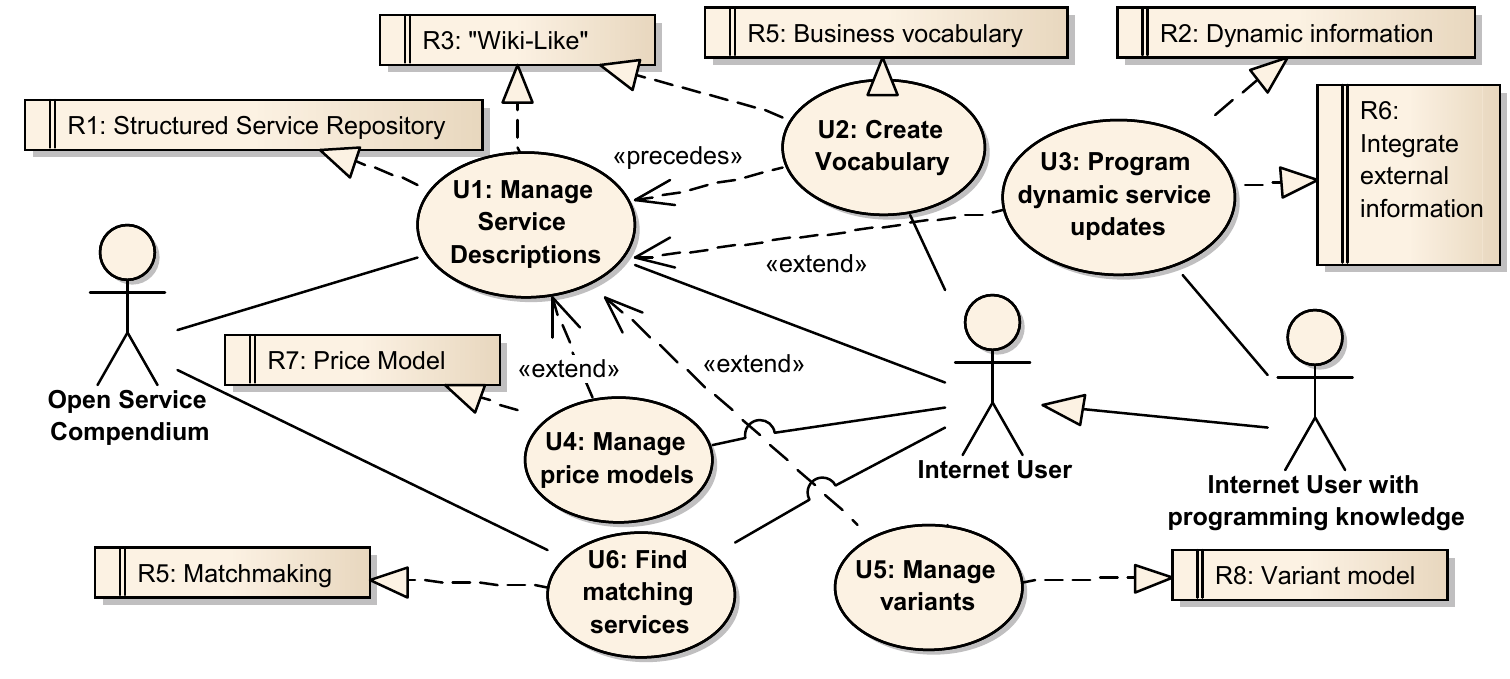}
\caption{OSC Use Cases}
\label{fig:use-case-model}
\end{figure}

This chapter derives the OSC use-cases which address the business challenges and their requirements illustrated in the preceding chapter. The use-cases are numbered U1 to U6 and are described in detail while Figure \ref{fig:use-case-model} provides the UML use-case diagram for easy reference.

\setlength{\belowcaptionskip}{-15pt}

\textbf{U1: Manage Service Descriptions.} The OSC should provide functionality for Internet users to manage service descriptions, i.e, to create, show, edit, and delete structured service descriptions to provide a \textit{structured service repository} (R1). As the OSC is "wiki-like" (R3), the use-case should allow a comprehensible, text-based language for service descriptions.

\textbf{U2: Create Vocabulary.} The \textit{business-pertinent vocabulary} (R5) provides the structure for service descriptions managed by Use Case U1. Thus, the OSC should provide Internet users the functionality to create such a vocabulary. To preserve the "wiki-like" characteristic of the OSC, the vocabulary definition has to be based on a comprehensible, text-based language. In order to support companies with their assessment, the vocabulary should also contain additional information on \textit{why} a certain property is important for service selection.

\textbf{U3: Program dynamic service information updates.} To \textit{integrate external information} (R6) and to also allow service knowledge to be \textit{dynamically updated} (R2), the OSC should give Internet users with programming knowledge the ability to implement dynamic service information updates. For reduced complexity, dynamic and static parts of a service descriptions should be integrated tightly, e.g., by having both in the same document. Dynamically updated service descriptions can lower the authoring effort considerably.

\textbf{U4 \& U5: Manage price models and service variants.} In order to handle complex price and variant structures, the OSC should manage \textit{price models} (R7) as well as \textit{variants} (R8) of services including creating, showing, editing and deleting these models, as well as presenting a price calculator, a variants overview, and using the price and feature model within service comparisons. The OSC should evaluate these models when finding services, i.e., it should find the respective variant of a service matching the search request as well as calculating the prices of those variants. This model should be integrated into the description language in order to reduce complexity and to allow features to be linked to their impact on the price of service consumption.

\textbf{U6: Find matching services.} All of the previous use-cases culminate in the pivotal use-case of enabling Internet users to \textit{find matching services to their requirements} (R5). For this, end-users should be able to define their requirements in an interactive fashion. The OSC should then evaluate services, price, and feature models to present matching services. In addition to basic selection, the OSC should also contain a matchmaking component in order to rank services and provide a more comprehensive selection result.

\section{Open Service Compendium Architecture}
\label{sec:architecture}

This chapter presents the Open Service Compendium architecture which is designed to implement the OSC use cases by providing insight on three layers: the conceptual architecture, its implementation, as well as its future expansion.

\begin{figure}
\centering
\includegraphics[width=8cm, clip=true, trim=3.9cm 19.4cm 4.1cm 0.7cm]{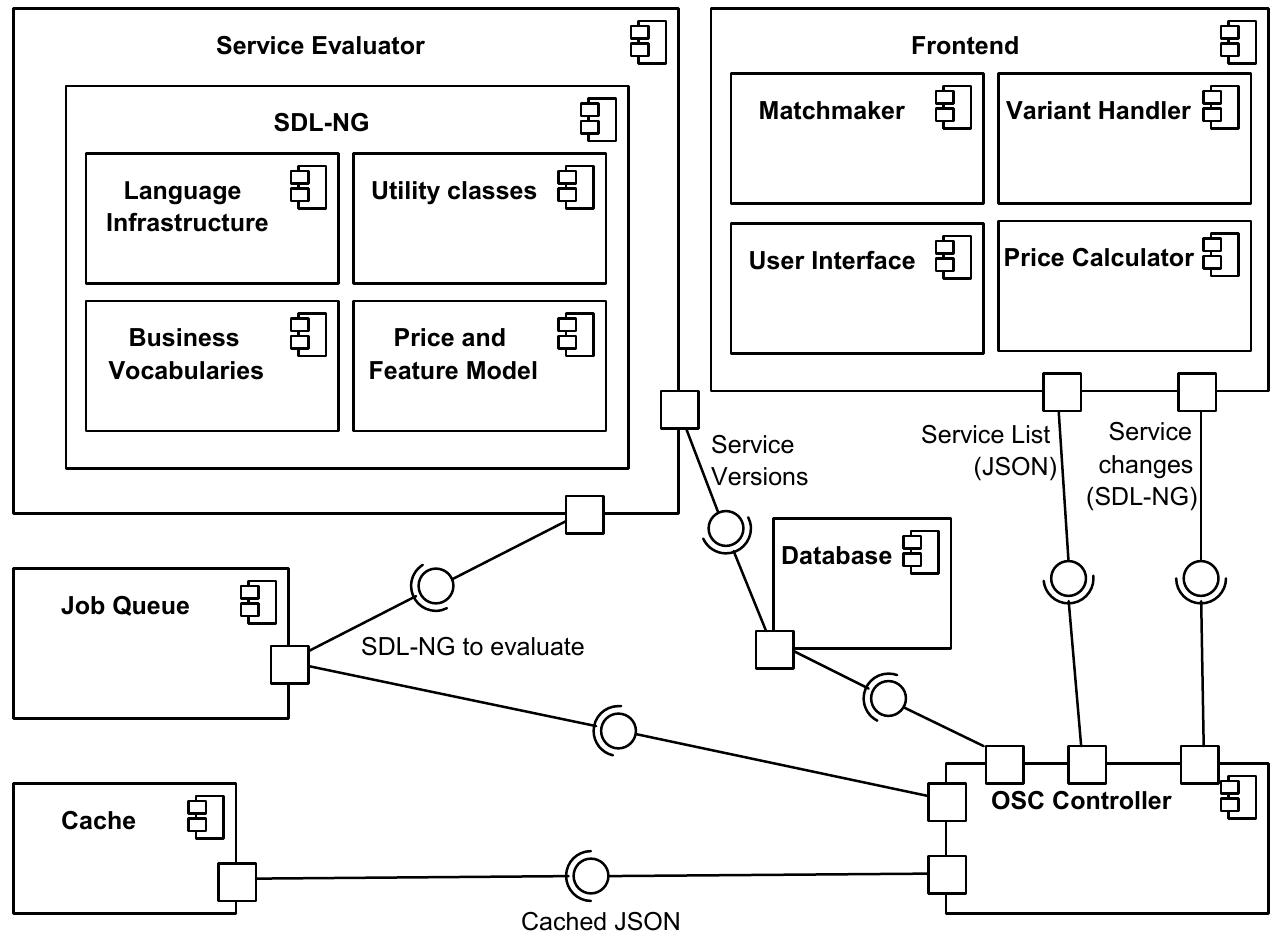}
\caption{Conceptual architecture}
\label{fig:architecture}
\end{figure}

\textbf{The conceptual architecture.} Figure \ref{fig:architecture} presents the conceptual architecture in form of a UML component diagram. There are six main components, which are described in the following paragraphs.

The pivotal \textbf{OSC controller} is responsible for providing a service list as JSON to the \textit{Frontend}, either by querying the \textit{Database} or using the \textit{Cache}. Changes to service descriptions are recieved from the \textit{Frontend} in SDL-NG form and submitted to the \textit{Job Queue} for subsequent evaluation. The \textbf{Job Queue} can additionally save the evaluation status, e.g., in cases of erroneous SDL-NG documents. The \textbf{Service Evaluator} takes SDL-NG documents from the \textit{Job Queue}, executes them in a specially secured container to prevent malicious code execution, and writes the resulting service descriptions to the \textit{database}. The \textbf{Frontend} provides the end-user interface for the use cases U1 through U6. It is designed as a JavaScript "Single Page Application" and is responsible for querying the OSC controller for JSON service descriptions, service matchmaking, service variant handling, price calculation, as well as submitting new and modified service descriptions to the OSC controller. A number of factors led to our adoption of the emerging "Single Page Application" style, instead of having the OSC controller render all the pages: the \textit{back end simplicity}, the \textit{swiftness} of the user interface, as well as the \textit{decoupling} of back end and front end. As our architecture is well decoupled, each component can be scaled independently.

\textbf{Implementation.}
\label{sec:architecture-implementation}
The OSC prototype can be accessed online \footnote{\url{h
ttp://www.open-service-compedium.org}} and its current source code can be found on Github\footnote{\url{https://github.com/TU-Berlin-SNET/open-service-compendium}}. While we describe the final implementation, some of the components are currently under development and not publicly available for testing.

We used the brokering component of the TRESOR project as the base for creating the OSC controller, a Ruby on Rails application offering RESTful APIs for creating, querying, updating, and deleting services stored in a MongoDB database. The database holds service documents which contain a persisted representation of the executed ruby service description as well as meta-information, such as the execution timestamp. The Job Queue uses a Redis key-value store managed by the Resque Ruby library. For caching, we rely on the Rails-default \texttt{ActiveSupport::Cache} infrastructure, as it is highly flexible in its use of different caching stores, for example, in-memory, plain files, MemCache, or Redis. The service evaluator is a regular Ruby process using the SDL-NG to evaluate the service descriptions in a secure context and persisting the resulting service descriptions in the database. The SDL-NG contains a language infrastructure (e.g., a type system and exporters), utility classes for scraping websites, business vocabularies (e.g, for cloud storage and IaaS offerings), as well as a price and feature model. As SDL-NG descriptions can scrape websites with changing content, e.g., Amazon Spot Market prices, the service evaluator can be instructed to regularily check the description for changes and update the database records accordingly. This functionality can handle dynamic aspects of cloud systems, e.g., pricing and capability changes. Historical service records can be retrieved using the OSC controller, for example, to detect modifications of certain services and to make predictions about future changes (e.g., price drops). Our previous publication \cite{SK14} provides an in-depth explanation of the SDL-NG.

\setlength{\belowcaptionskip}{-15pt}
\begin{figure}
\vspace*{-0.5cm}
\begin{center}
\includegraphics[width=9cm]{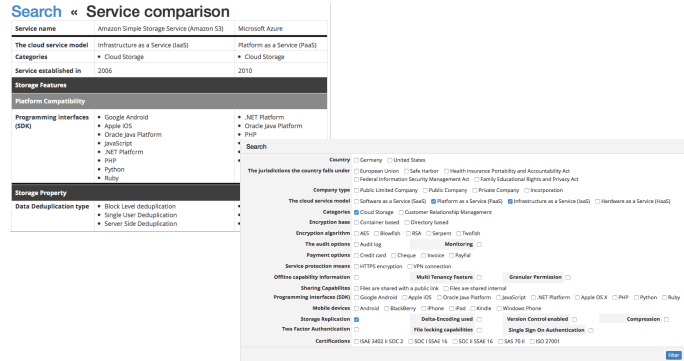}
\end{center}
\caption{User interface: faceted search, service comparison, and storage vocabulary}
\label{fig:screenshots}
\end{figure}

The frontend is built and assembled using a Grunt JavaScript workflow. It is based on AngularJS and a number of additional Javascript and CSS libraries, e.g., Angular UI Router, Twitter Bootstrap, Less, SASS, and CoffeeScript. It integrates a Java applet containing a constraint programming matchmaker using the Choco Solver\footnote{\url{http://www.emn.fr/z-info/choco-solver/}} to implement different constraint models for different matchmaking functionality: \textit{discrete value matching} with hard and soft constraints, \textit{interval matching} for negative and positive tendencies, and \textit{matching of feature lists}. Our previous work \cite{ZSK15} includes a detailed description of the constraint models and their implementation. Figure \ref{fig:screenshots} shows some example screenshots of the user interface. In general, there are three main views for users to realize service \textit{discovery}, \textit{assessment}, and \textit{selection}. First, a list of all services in a specific category (e.g., cloud storage and IaaS solutions) allows users to \textit{discover} all available services. Users can use a faceted search to filter the list based on their selection criteria, such as company jurisdiction, payment options, and certifications. For \textit{assessment} there is a "detail" view, where users can see the whole service description, including extensive documentation about all properties and their meaning, as well as a comparison view. To support their final \textit{selection}, users can employ the matchmaking component for ranking services based on user defined constraints and getting the "best" service for their needs.

\textbf{Future expansion.} Through eventual OSC advancements, we foresee some prospective components: first of all, there has to be some kind of basic \textit{user management} to protect descriptions from vandalism or malicious editing. Secondly, to strengthen the usefulness of the OSC, additional external information sources should be included and managed by OSC components, e.g., external service reviews, user ratings, and benchmarking data. At last, the RDF/OWL export capabilities of the \textit{Service Description Language} could be used to implement a \textit{Semantic Data Store} component in order to publish an OSC dataset in the \textit{Linked Open Data Cloud}. This has two main goals: raising the business relevance of related approaches by offering semantic descriptions of real-world services as well as enabling the OSC to benefit from advanced functionality, such as machine learning and reasoning.

\section{Evaluation}
\label{sec:evaluation}

This chapter presents the evaluation of the OSC architecture and its implementation: \textit{analytical}, comparing it to the set of general requirements, \textit{experimental}, gathering knowledge from using it in practice, as well as \textit{empirical}, carrying out interviews and surveys.

\textbf{Analytical Evaluation.} Carrying out the analytical evaluation is straightforward and highlights the fitness of the OSC to cover all enumerated requirements: we have created a \textit{structured service repository} \textbf{(R1)} using a comprehensive Service Description Language. \textit{Dynamic information} \textbf{(R2)} can be continuously integrated by frequently running the \textit{Service Evaluator}, as explained in Chapter \ref{sec:architecture-implementation}. As we have chosen a simple to use textual DSL, the OSC becomes a \textit{"wiki-like"} information system \textbf{(R3)}, which was also highlighted in our recent publication \cite{SK14}. We integrated a \textit{service matchmaker} \textbf{(R4)} to match requirements with structured service knowledge. The business and category-specific vocabularies contain empirically determined \textit{selection criteria}, which are pertinent to businesses \textbf{(R5)}, which is highlighted by our empirical evaluation. Internet Users with programming knowledge can \textit{integrate external information} \textbf{(R6)} using the Utility Classes of the SDL, as exemplified in the service description examples\footnote{\url{https://github.com/TU-Berlin-SNET/sdl-ng/tree/master/examples/services}}. The SDL-NG contains an additional \textit{price model} \textbf{(R7)} as well as a model to capture service \textit{variants} \textbf{(R8)}.

For \textbf{experimental evaluation}, we have applied the OSC in practice to validate its functionality by implementing an automated test suite, as well as manual usage. So far, the OSC functions as specified. For example, users can have a look at the vocabulary "cheat sheet" to get an overview of all properties and types, use a code editor to create service descriptions, view the output of the service evaluation, search for services, and compare them.

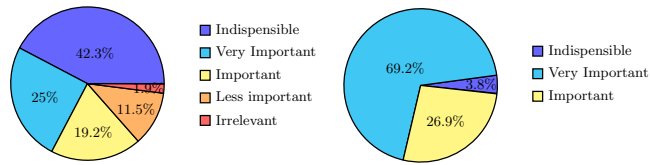
\begin{figure}
\vspace*{-0.5cm}
\begin{center}
\begin{tabular}{c c}
\scalebox{0.6}{
\begin{tikzpicture}
\pie[text=legend, radius=1.7]{42.3/Indispensible, 25/Very Important, 19.2/Important, 11.5/Less important, 1.9/Irrelevant}
\end{tikzpicture}
}

&

\scalebox{0.6}{
\begin{tikzpicture}
\pie[text=legend, rotate=-6, radius=1.7]{3.8/Indispensible, 69.2/Very\ Important, 26.9/Important}
\end{tikzpicture}
}

\end{tabular}
\end{center}
\caption{Survey results for business (left) and cloud storage vocabulary (right)}
\label{fig:piechart}

\end{figure}

Additional \textbf{empirical evaluation} was carried out for the OSC components \textit{service description language}, \textit{business vocabulary}, and the \textit{cloud storage vocabulary}. The \textit{service description language} was presented to a group of experts from other \anonymize{Trusted Cloud} projects having been involved in related research, e.g., the USDL \cite{OBKH13}. A group discussion about the relevance of the OSC for the field resulted in the following statements: the textual DSL is a major simplification in describing services, especially in comparison to the USDL and semantic approaches. Deriving the vocabulary from empiric research strengthens the usefulness of the DSL in practical contexts. The utility value of a central repository with matchmaking capabilities was regarded as very high. An expert group evaluated the \textit{business vocabulary} with respect to its relevance for service selection. Participants were one publication author, the CIO, and two IT project managers of a large German health center. They had to come up with a mutual importance rating of the individual criteria on a 5-step scale from "indispensible" (1) to "irrelevant" (5). The left pie chart in Figure \ref{fig:piechart} groups the categories by their rated importance: 86.5\% of the 52 criteria were rated important and higher, while only 13.5\% were rated less important or irrelevant. Evaluation of an intermediate version of the \textit{Cloud Storage Vocabulary} was performed using an online questionnaire in which participants rated the importance of the 27 criteria for their selection of a cloud storage service. The respondents were mostly students of computer science and related fields of study, providing valuable insight into the usefulness for generic Internet users, as most other people involved in our research are either professionals or academics. Of 35 respondents, 18 (51.4\%) completed the questionnaire. The right pie chart in Figure \ref{fig:piechart} shows the distribution of the average importance of all criteria, which is grouped into $1-1.5$ (indispensable), $1.5-2.5$ (very important), and $2.5-3.5$ (important). These results show that the generic OSC business vocabulary is able to capture some of the most important selection criteria of this specific client. For generalization, we will conduct an extensive questionnaire in the future and adjust the business vocabulary according to its results. The results promise a high relevance of the vocabulary for Internet users, yet the absence of less important and irrelevant criteria could also point at the inability of our respondents to differentiate the importance of their criteria. The low completion rate could imply that people either could not understand the criteria or did not know their cloud provider selection process.

\setlength{\belowcaptionskip}{-15pt}

\section{Conclusion}
\label{sec:conclusion}

We have delineated the challenges in discovery, assessment, and selection of cloud services and revealed the failure of both academic and commercial approaches to address these challenges properly. By using real-world requirements as the basis for the OSC use cases as well as designing a modern solution architecture, we hope to create a practical, mature, simple and usable information system. Preliminary evaluation results are promising and we are looking forward to presenting and discussing our contribution with practitioners and researchers at GECON. We also hope to maximize the impact of the OSC by creating a "wiki-like" information system and publishing it as free and open source (FOSS) software. In the future, we will use the OSC as a basis to tackle upcoming challenges within federated, multi-cloud environments and the Intercloud in the \anonymize{CYCLONE} project.

\section*{Acknowledgments}

This work is supported by the Horizon 2020 Innovation Action CYCLONE, funded by the European Commission through grant number 644925.

\bibliographystyle{splncs03}
\bibliography{gecon2015,ilke}

\end{document}